\documentclass[pdflatex,sn-mathphys-num]{sn-jnl}

\usepackage{graphicx}%
\usepackage{multirow}%
\usepackage{amsmath,amssymb,amsfonts}%
\usepackage{amsthm}%
\usepackage{mathrsfs}%
\usepackage[title]{appendix}%
\usepackage{xcolor}%
\usepackage{textcomp}%
\usepackage{manyfoot}%
\usepackage{booktabs}%
\usepackage{array}%
\usepackage{colortbl}%
\usepackage{algorithm}%
\usepackage{algorithmicx}%
\usepackage{algpseudocode}%
\usepackage{listings}%
\usepackage[utf8]{inputenc}
\usepackage[T1]{fontenc}
\usepackage{CJKutf8}
\usepackage{wrapfig}
\usepackage{float}

\usepackage{iftex}
\ifPDFTeX
  \DeclareUnicodeCharacter{202F}{\,}
  \DeclareUnicodeCharacter{2011}{-}
  \DeclareUnicodeCharacter{2212}{-}
\else
  \usepackage{newunicodechar}
  \newunicodechar{ }{\,}
  \newunicodechar{‑}{-}
  \newunicodechar{−}{-}
\fi

\theoremstyle{thmstyleone}%
\theoremstyle{thmstyletwo}%

\theoremstyle{thmstylethree}%

\begin{document}

\title[PSD2Code: Automated Front-End Code Generation]{PSD2Code: Automated Front-End Code Generation from Design Files via Multimodal Large Language Models}


\author{\fnm{Yongxi} \sur{Chen}}\email{71265901021@stu.ecnu.edu.cn}

\author*{\fnm{Lei} \sur{Chen*}}\email{lchen@cs.ecnu.edu.cn}

\affil{\orgdiv{School of Computer Science and Technology}, \orgname{East China Normal University}, \orgaddress{\street{3663 North Zhongshan Road}, \city{Shanghai}, \postcode{200062}, \state{Shanghai}, \country{China}}}


\abstract{Design-to-code generation has emerged as a promising approach to bridge the gap between design prototypes and deployable frontend code. However, existing methods often suffer from structural inconsistencies, asset misalignment, and limited production readiness. This paper presents PSD2Code, a novel multimodal approach that leverages PSD file parsing and asset alignment to generate production-ready React+SCSS code. Our method introduces a Parse–Align–Generate pipeline that extracts hierarchical structures, layer properties, and metadata from PSD files, providing large language models with precise spatial relationships and semantic groupings for frontend code generation. The system employs a constraint-based alignment strategy that ensures consistency between generated elements and design resources, while a structured prompt construction enhances controllability and code quality. Comprehensive evaluation demonstrates significant improvements over existing methods across multiple metrics including code similarity, visual fidelity, and production readiness. The method exhibits strong model independence across different large language models, validating the effectiveness of integrating structured design information with multimodal large language models for industrial-grade code generation, marking an important step toward design-driven automated frontend development.}

\keywords{Large Language Model, Front-End Code, Multimodal}



\maketitle

\section{Introduction}\label{sec1}
Graphical user interfaces (GUIs) are central to how users perceive and interact with modern applications. For web products in particular, GUI quality directly affects usability, user experience, and business growth, making pixel-accurate layout, visual consistency, and semantic structure critical to product success. However, implementing production-grade GUIs remains labor-intensive: front-end engineers repeatedly translate design artifacts into code, refine layout details, and reconcile assets and styles to meet engineering conventions. This repetitive, time-consuming workflow motivates research on automatic front-end code generation.

Existing approaches to Design-to-Code can be summarized along a development trajectory from traditional rule/heuristic pipelines, to deep learning methods, and more recently, large multimodal language models. For completeness, when discussing text/code similarity baselines, we also adopt terminology from classic NLP metrics such as BLEU \cite{papineni2002bleu}.

(1) Traditional pipelines rely on templates, rules, and computer-vision heuristics to map recognized components to code. Representative systems include Microsoft's Sketch2Code, which extracts elements from sketches or screenshots and applies rule-based templates to generate HTML/CSS \cite{microsoft2018sketch2code}. Industrial platforms such as Imgcook and CodeFun emerged later, providing engineering-oriented tooling for design-to-code translation \cite{imgcook,codefun}. These pipelines work well in standardized scenarios but struggle with complex layouts, deep hierarchies, and maintainable component structures.

(2) Deep learning methods learn to translate visual inputs into code or structured representations. Early work such as Pix2Code maps screenshots to DSL/code using sequence models \cite{beltramelli2017pix2code}. ReDraw leverages large mined UI corpora to infer component hierarchies from screenshots and retrieve stylistically similar components for code generation \cite{moran2018redraw}. Subsequent research reconstructs GUI skeletons from design images to bootstrap implementations \cite{chen2019}. While these methods improve recognition and layout recovery, they still face challenges in preserving precise hierarchy, ensuring resource traceability, and generating production-ready code.

(3) Large multimodal language models (LLMs) represent the recent trend, enabling instruction-driven code generation that better integrates visual cues with code priors. General-purpose models (e.g., GPT-4) demonstrate strong code generation and reasoning capabilities \cite{openai2023gpt4}. However, purely vision/text-driven pipelines can still suffer from hierarchy drift and resource mismatches without explicit design metadata.

Despite the strong capabilities of large-model-based UI-to-code generation methods in end-to-end automation and complex layout understanding, significant challenges remain in industrial production settings. Purely vision- or text-driven generation methods can still exhibit hierarchical or layout deviations when handling complex nested structures. Although model outputs have improved in quality through hierarchy-aware and self-correcting mechanisms, potential issues remain in style consistency, semantic naming, and component boundaries, making it difficult to fully comply with existing engineering systems and standards. Furthermore, in industrial deployment, challenges persist in material alignment and resource management, such as robustly mapping coordinates, hierarchy, and properties from design files to the final code to ensure rendering consistency with the design. These issues indicate that, despite the potential of large models in UI-to-code generation, additional integration with structured design file information and output verification is necessary to achieve production-ready, industrial-grade code.

Motivated by these limitations, we advocate a PSD-first, metadata-aware UI-to-code approach that parses hierarchical structures, coordinates, and asset references directly from design files, and uses these structured priors to constrain multimodal generation. This design-aware strategy aims to reduce repetitive manual work while improving visual fidelity, semantic structure, and production readiness of the generated front-end code. We conducted a systematic evaluation on a real-world PSD design dataset, primarily consisting of large-scale game event pages, landing pages on both PC and mobile platforms, and pop-up components. The dataset contains approximately 100 samples with project-level deduplication and strict training/validation/test splits (70\%/15\%/15\%) to prevent data leakage. Each sample includes original PSD files, corresponding screenshots, asset files (3-15 assets per sample), ground-truth React+SCSS implementations, and rendered outputs. Evaluation was performed across three core dimensions: code similarity (CodeBLEU, CodeBERT vector similarity), visual similarity (SSIM, PSNR, MSE), and human evaluation (subjective scores on readability, maintainability, and visual consistency), with additional layout consistency metrics (approximate mAP, APs/APm/APl). Baselines include screenshot-to-code (GPT-4V), CodeFun, and pix2code. Our method achieves approximately 37.2\% increase in visual similarity (SSIM) and 5.6\% increase in PSNR, while maintaining high success rates in code generation (98.4\%), rendering (99.2\%), and resource integration (94.2\%). The method exhibits strong model independence across four different large language models with performance variation within 5.2\%, demonstrating clear overall superiority over existing approaches.

\begin{figure}[H]
\centering
\includegraphics[width=\linewidth,height=0.3\textheight,keepaspectratio]{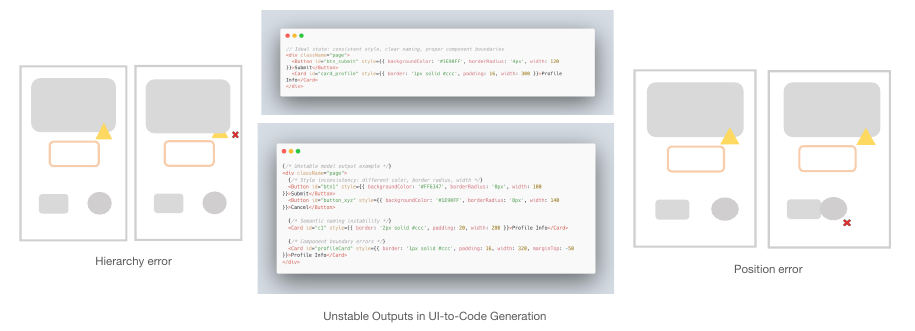}
\caption{Limitations of large-model-based UI-to-code generation in industrial settings.}
\label{fig:llm-limitations}
\end{figure}

\begin{figure}[H]
\centering
\includegraphics[width=\linewidth]{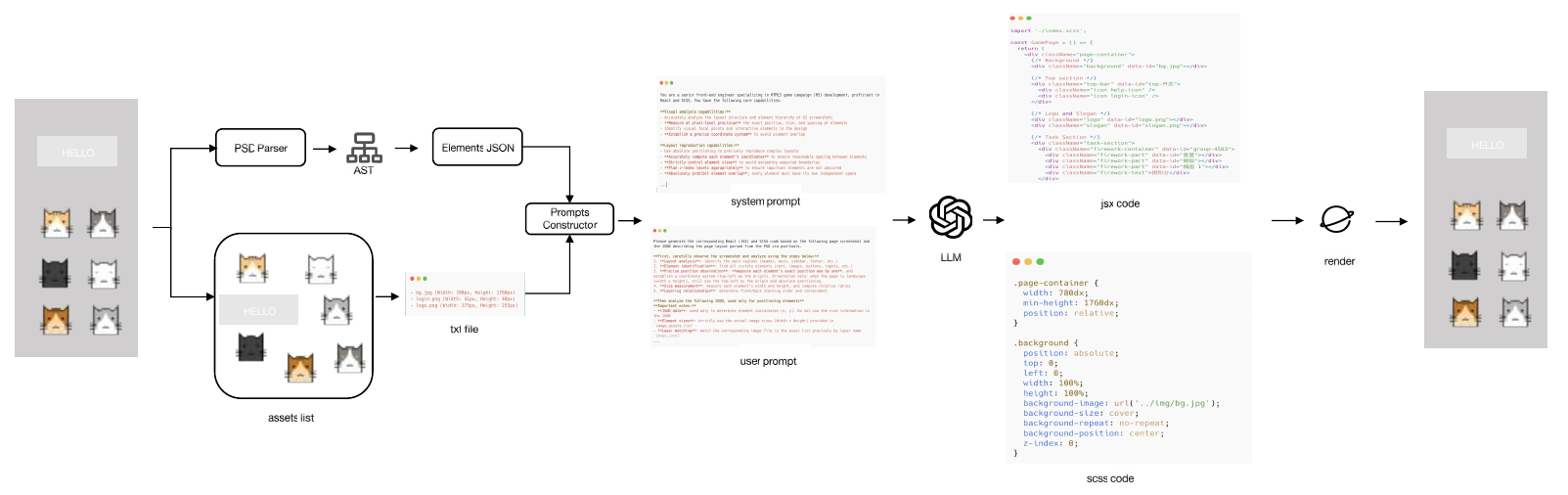}
\caption{System overview of the proposed Parse-Align-Generate framework.}
\label{fig:overview}
\end{figure}

In summary, we adopt a closed-loop "Parse–Align–Generate–Validate" framework to operationalize the above design-aware strategy (see Fig.~\ref{fig:overview}); detailed methodology is presented in Section~\ref{sec:methods}.

Our main contributions are as follows:

\begin{itemize}
\item We propose a closed-loop "Parse–Align–Generate–Validate" framework that explicitly incorporates structured information from design files (e.g., coordinates, hierarchy, and asset references) into the generation process. This framework constrains large models with hard conditions to output React + SCSS code that is directly usable for development, significantly enhancing industrial applicability.

\item We construct a comprehensive PSD design dataset covering multiple scenarios, including game landing pages, lottery pages, showcase pages, and pop-ups, across both PC and mobile platforms. The dataset provides parsed JSON, asset lists, ground-truth JSX/SCSS, and rendered screenshots, enabling end-to-end evaluation and reproducible experimental pipelines.

\item Our method consistently outperforms baseline methods (screenshot-to-code, CodeFun, pix2code) across code similarity, visual similarity (SSIM, PSNR, MSE), and human evaluation (readability, maintainability, visual consistency). It also achieves superior layout consistency (approximate mAP, APs/APm/APl), demonstrating its effectiveness in generating production-ready code.
\end{itemize}

\section{Related Work}\label{sec:related}

In large-scale frontend industrial development, collaboration between designers and engineers is crucial. Designers are responsible for producing UI prototypes, while frontend engineers code based on these prototypes. Due to the existence of numerous similar elements and functions in frontend applications, this means there is a large amount of repetitive and meaningless work in the frontend development process. To improve efficiency, frontend automatic code generation technology has emerged. Compared to code generation based solely on visual representations, the path from design prototypes to code, with the support of multimodal information, better aligns with the needs of industrial development.

\subsection{Technological Evolution of Design-to-Code Generation}\label{subsec:evolution}

In recent years, with the rise of large language models and multimodal approaches, this field has achieved significant progress, making frontend project automation a promising area. The evolution can be categorized into three main phases:

\subsubsection{Traditional Rule-Based and Template Methods}

Early approaches to design-to-code generation relied heavily on predefined rules, templates, and computer vision heuristics. Design prototypes are typically constructed by professional designers, and compared to pure visual representations, their acquisition process is more complex and costly. Most prototype-to-code methods originate from professional design platforms or large internet companies. For example, commercial platforms like Figma to Code and Builder.io parse structured metadata in design prototypes through predefined rules and heuristic algorithms, then automatically generate corresponding frontend code. However, due to the lack of deep understanding of precise positioning and hierarchical structure in design files, the generated code often has deviations in element positioning and layout restoration, unable to generate code that is directly usable in engineering. Additionally, tools like Imgcook and CodeFun analyze interface structure through computer vision technology, convert recognition results into intermediate representation languages, and then generate multi-framework compatible code through rule engines. Although such methods have certain advantages when handling batch requirements, they still have problems such as inaccurate element recognition and insufficient understanding of visual grouping when handling complex designs, often requiring manual intervention to achieve expected results.

\subsubsection{Deep Learning-Based Approaches}

The second phase saw the emergence of deep learning methods that learn to translate visual inputs into code or structured representations. Early work such as Pix2Code maps screenshots to DSL/code using sequence models \cite{beltramelli2017pix2code}. ReDraw leverages large mined UI corpora to infer component hierarchies from screenshots and retrieve stylistically similar components for code generation \cite{moran2018redraw}. Subsequent research reconstructs GUI skeletons from design images to bootstrap implementations \cite{chen2019}. Liu et al. proposed Prototype2Code, an end-to-end method from UI design prototypes to frontend code \cite{liu2024}. Zhang et al. proposed a divide-and-conquer strategy, first cutting screenshots and then further generating UI code, emphasizing the importance of component-level understanding \cite{zhang2024}. Chen et al. conducted a comprehensive evaluation of design-to-code tools, providing valuable insights into the current state of research and identifying key challenges \cite{chen2024evaluation}. While these methods improve recognition and layout recovery, they still face challenges in preserving precise hierarchy, ensuring resource traceability, and generating production-ready code.

\subsubsection{Large Language Model}

The year 2023 marked a significant shift toward large language model approaches. General-purpose LLMs for code such as Codex, AlphaCode, and Code Llama established strong baselines for program synthesis and code generation \cite{chen2021codex,li2022alphacode,roziere2023codellama}. Reasoning and tool-use techniques further improved reliability and controllability, including chain-of-thought prompting, self-consistency, tool usage, and retrieval-augmented generation \cite{wei2022cot,wang2022selfconsistency,schick2023toolformer,lewis2020rag}. On the vision side, representation learning via CLIP enabled robust multimodal grounding that benefits UI understanding \cite{radford2021clip}. Zhu et al. proposed a visual critic-based UI-to-code reverse generation method, based on the non-differentiable nature of rendering, eliminating the need for rendering and thus solving the training data generation problem \cite{zhu2023}. With large language models (LLMs) demonstrating excellent performance in understanding requirements and code generation, platforms like Anima have introduced generative AI technology, enabling developers to generate code through natural language descriptions of requirements. However, these LLMs' performance in layout reasoning has not yet met expectations, and the generated code still requires manual optimization in terms of responsiveness and maintainability.

Recent advances in vision-language models are revolutionizing the UI-to-code generation field. Wang et al. proposed Web2Code, a large-scale webpage-to-code dataset and evaluation framework specifically designed for multimodal large language models, highlighting the importance of high-quality training data in advancing this field \cite{wang2024web2code}. Yang et al. proposed the WebSight dataset for converting webpage screenshots to HTML code \cite{yang2024}; Wang et al. developed InternLM-XComposer-2.5, a general-purpose vision-language model supporting long-context input and output, particularly suitable for complex UI design \cite{wang2024internlm}. In terms of benchmarks and evaluation, Wei et al. proposed RoboCodeX for multimodal code generation in robot behavior synthesis, demonstrating the broad applicability of multimodal approaches \cite{wei2024}. Ma et al. proposed MMCode, a comprehensive benchmark for evaluating multimodal large language models' performance in code generation tasks involving visually rich programming problems \cite{ma2024}.

Recent works continue to explore hierarchical generation and design-metadata-aware pipelines that first construct high-level structure before emitting detailed code and styles, aiming to improve both fidelity and maintainability. Benchmarks and datasets for UI understanding and layout analysis (e.g., RICO and PubLayNet) have also supported progress in this area \cite{deka2017rico,zhong2019publaynet}.

The conversion from sketches and prototypes to code is another important direction in this field. Sun et al. evaluated the performance of vision-language models in interactive webpage design prototypes \cite{sun2024}, while Lin et al. proposed UI layout generation methods based on UI syntax constraints, emphasizing the importance of structural constraints in code generation \cite{lin2023}. Li et al. explored generating HTML code from handwritten images, demonstrating the diversity of visual input methods \cite{li2023}. This research complements our PSD-based approach, showing alternative ways to capture design intent.

Although most existing research focuses on generation based on screenshots or sketches, few involve the rich information in design files (such as PSD). Adobe's Design-to-Code plugin and tools like Zeplin provide commercial solutions, but academic research in this field remains limited. Our research directly parses PSD files to extract precise positioning and style information. Compared to screenshot-based methods, this approach can provide more accurate spatial relationships and represents a contribution by directly utilizing design metadata rather than relying solely on visual interpretation.

\subsection{Limitations of Existing Methods and Our Approach}\label{subsec:limitations}

Existing methods are insufficient in element-level precise positioning and hierarchy restoration: most only infer based on screenshots, lacking reliable coordinates and hierarchical structures obtained from design file parsing, leading to positioning drift and structural ambiguity. Insufficient consideration of responsive and cross-form adaptation: generally generating static pages without systematically handling landscape/portrait orientation, viewport changes, and adaptive backgrounds and other engineering requirements. Limited code maintainability and engineering usability: many methods output DSL or loose HTML/CSS, lacking modular styles, semantic structure, and asset management, making them difficult to use directly in production. Insufficient evaluation in real scenarios: evaluation is mostly based on synthetic or simplified data, ignoring practical constraints such as high proportions of image asset references, complex overlays, and resource paths in business projects.

Our PSD2Code system addresses some of these limitations: adopting a PSD-first approach, unlike screenshot-based methods, we directly parse PSD files to extract precise layout information, asset references, and layer hierarchies, thereby obtaining design metadata that visual methods cannot capture; implementing multimodal integration, combining PSD parsing results, image assets, and screenshot information to jointly input structured and visual information into language models; establishing a comprehensive evaluation system, including code similarity metrics (CodeBLEU, CodeBERT), visual similarity metrics (SSIM, PSNR, MSE), and layout detection metrics (mAP, APs, APm, APl), forming an overall evaluation system; evaluating based on real datasets, using real UI designs rather than synthetic datasets, closer to actual production environment needs; having resource-aware generation capabilities, the system has awareness of design assets and resources, ensuring generated code can correctly reference and utilize existing resources.

By combining structured design information (PSD) with visual understanding, our method fully leverages the advantages of design file analysis and computer vision technology. This approach bridges traditional design tool workflows with modern AI-driven code generation, providing practical value for real development scenarios.

\section{Methodology}\label{sec:methods}

\subsection{Overall Framework Design}

Our methodology adopts a structured, three-stage pipeline that transforms PSD design files into production-ready React+SCSS code through systematic parsing, alignment, and generation. The framework is designed to address the fundamental challenges in design-to-code generation: maintaining precise spatial relationships, preserving hierarchical structure, ensuring asset consistency, and producing executable, maintainable code.

Our method follows a three-stage parsing-alignment-generation pipeline. In the parsing stage, the system accurately extracts effective layer information and absolute positioning data from PSD files to construct a structured design blueprint. During the alignment phase, this blueprint undergoes rigorous validation to establish one-to-one mapping between resource references and actual engineering assets, ensuring precise path and dimension matching. Finally, in the generation stage, multimodal large language models are leveraged with specifically engineered prompts that incorporate the comprehensively aligned asset information—including validated resource paths and dimensional parameters—alongside design screenshots to produce production-ready React+SCSS code.

This design philosophy prioritizes structured information over pure visual interpretation and constraint-based generation over free-form synthesis. By explicitly encoding design metadata and enforcing hard constraints during generation, our approach achieves superior accuracy in spatial positioning, hierarchical preservation, and resource management compared to existing screenshot-based methods.

\begin{table}[!htbp]
\centering
\caption{PSD2Code Algorithm: Parse-Align-Generate Pipeline}
\label{tab:algorithm}
\begin{tabular}{p{0.1\linewidth}>{\raggedright\arraybackslash}p{0.8\linewidth}}
\hline
\textbf{Step} & \textbf{Description} \\
\hline
& \textbf{Input:} PSD file $P$, Asset directory $A$ \\
& \textbf{Output:} React+SCSS code $C$ \\
\hline
& \textbf{Stage 1: Parse} \\
1 & Load PSD file $P$ and construct layer tree $L$ \\
2 & Apply multi-layer filtering: remove hidden/transparent layers, minimal areas, and system-default named elements \\
3 & Normalize coordinates: clip positioning data to page pixel coordinate system \\
4 & Classify layers: $type \in \{\text{container}, \text{text}, \text{image}\}$ (priority: container $\to$ text $\to$ image) \\
5 & Generate design.json: $\{dimensions, elements[], assets[]\}$ \\
6 & Optimize hierarchy: remove empty containers, control depth (MAX\_DEPTH=6) \\
\hline
& \textbf{Stage 2: Align} \\
7 & Extract asset list from $A$ with actual dimensions \\
8 & Match parsed elements to asset resources \\
9 & Enforce size constraints: $size \leftarrow asset\_dimensions$ \\
10 & Validate resource traceability and file paths \\
\hline
& \textbf{Stage 3: Generate} \\
11 & Construct multimodal prompt: $\{design.json, assets\}$ \\
12 & Apply hard constraints: coordinates, hierarchy, sizes \\
13 & Generate React+SCSS code using LLM \\
14 & Render and validate output quality \\
\hline
& \textbf{Return:} Production-ready code $C$ \\
\hline
\end{tabular}
\end{table}

\subsection{PSD Parsing: From Layered Design to Structured Representation}\label{subsec:pipeline}

In this study, PSD parsing treats screenshots as pixel-level anchors and layered designs as semantic and hierarchical priors, processing only PSD inputs that conform to specifications to ensure a stable and reproducible parsing process. The parsing workflow traverses the layer tree layer by layer and applies visibility and quality filtering during preprocessing, automatically removing hidden, nearly transparent, extremely small, or invalidly named layers. All element coordinates are normalized to the page pixel coordinate system and strictly aligned with the accompanying screenshot.png, with out-of-bound or negative values clipped and normalized to ensure positions and sizes are non-negative integers within the page boundaries.

We have developed a classification methodology based on structural analysis and semantic inference to automatically categorize layers in PSD files into three fundamental types—container, text, or image—following a well-defined priority sequence that processes group layers first, then text layers, and finally image layers. For group layer processing, we implemented a dual-layer decision mechanism leveraging statistical features and pixel coverage ratios, where the algorithm begins by traversing sub-layers to quantify pixel-based layers, text layers, and nested groups while simultaneously computing the union bounding box of all pixel layers; by calculating the ratio of the united pixel area relative to the total group area, the system assesses the degree of visual element filling—when the coverage reaches $\geq$85\% with representative pixel candidates present, the group is folded into an image type adopting the optimal candidate's coordinates and dimensions, otherwise it is maintained as a container type with recursive sub-layer processing. To ensure classification precision, we established five constraints: mandatory inclusion of valid pixel layers meeting coverage thresholds; exclusion of groups containing text sub-layers; filtration of layers with structural identifiers such as masks or overlays; restriction of pixel candidates to $\leq$2 or elevation of coverage requirements to $\geq$95\%; and for layers with container semantics or subgroups, permitting folding only when satisfying both 95\% coverage and background-related keywords. For text layer classification, a resource-oriented decision strategy is employed where the system first performs exact matching to detect image resource files perfectly aligned with the layer name to effectively prevent redundant rendering of text and images, then conducts semantic matching via a preset resource library for short text ($\leq$3 characters) or specially formatted text ($\leq$10 characters), classifying matches as image type while retaining unmatched text layers with complete textual attributes as text type. Pixel-based layers are directly classified as image type with visual style features such as opacity extracted, and this comprehensive classification mechanism achieves precise differentiation between content images and structural containers through multi-dimensional verification, effectively handling text-image boundary cases to establish a reliable structured foundation for subsequent code generation.

The parsing results are exported as a unified design.json, with the top level recording page dimensions and the element tree. Each element includes id, name, position {x, y}, size {width, height}, and type (container/text/image), with missing fields filled with safe defaults to ensure stability for subsequent rendering and evaluation. Additionally, an assets list is constructed, recording each image resource's filename and actual width/height, while explicitly enforcing that element positioning relies solely on the JSON coordinates and sizes are determined by the true asset dimensions, avoiding size deviations caused by parsing noise.

Finally, consistency checks and scripted corrections are applied to remove empty containers and detect overlapping or out-of-bound elements, ensuring the parsed results maintain a clear hierarchy, semantic correctness, and traceable assets. This workflow provides a low-entropy, stable, and computable structured data foundation for subsequent prompt construction and React+SCSS code generation.

\subsection{Prompt Construction with PSD and Asset Integration: Low-Entropy Priors and Hard Constraint Injection}\label{subsec:generation}
To reduce the difficulty of learning structure and code directly from pixels and improve the industrial applicability of the generated code, we integrate PSD parsing results with the assets list to construct a multi-modal structured prompt, which serves as a strong constraint during the generation phase. The prompt consists of three parts:

\begin{itemize}
\item \textbf{Structural Prior Fragment}: Generated from design.json, including page dimensions, container hierarchy, element coordinates, and types, used only for positioning and hierarchical constraints, excluding fields that may introduce size noise.

\item \textbf{Asset Alignment Fragment}: Generated from the assets list, recording each image resource's filename and actual width/height, ensuring all resource dimensions are accurate and traceable, and explicitly stipulating that element sizes rely on assets rather than parsing results.

\item \textbf{Engineering Constraints and Output Specification Fragment}: Constrains the model to comply with engineering standards, including class naming conventions, executability, dual code block protocol (React+SCSS), absolute positioning, z-index planning, and prohibition of overlap or out-of-bound placement.
\end{itemize}

The prompt follows a three-part structure: system instruction, example, and user message. The system instruction defines the output format and engineering rules, prohibiting any additional explanatory text. The example demonstrates the mapping from design structural data to container hierarchy and reusable styles. The user message incorporates the structural prior and asset alignment fragments derived from PSD parsing, emphasizing the following constraints: positioning strictly adheres to JSON coordinates, element sizing depends on asset dimensions, images are referenced via background-image, text generation is exclusively applied to elements with type=text, and all top/left coordinates must be integers to ensure pixel-level precision. To enhance determinism and constraint strength, we apply noise suppression to the prompt by removing irrelevant dimensional and stylistic information, highlighting critical content, explicitly marking hard constraints, and providing element-by-element coordinate reminders for list and grid items, thereby ensuring strict spatial and resource-level control over the model's code generation process.
To enhance determinism and constraint strength, we apply noise suppression to the prompt by removing irrelevant size and style information, highlighting key information, explicitly marking hard constraints, and providing element-by-element coordinate reminders for list and grid items, ensuring strict spatial and resource-level control over the model.

\subsection{Large Model Generation: Structure-First React+SCSS Output with Executability Assurance}\label{subsec:evaluation}
During the generation phase, we adopt a structure-first, asset-aligned, and execution-controlled principle to guide the large model in producing React (JSX) and SCSS code that is directly usable for development. The strategy includes:

\begin{itemize}
\item \textbf{Generation Constraints and Output Specification}: Using design.json and assets as hard constraints, the model strictly follows element coordinates, hierarchy, and sizes; outputs adhere to the dual code block protocol (jsx and scss) without mixing or explanatory text; coordinates are integers for precise absolute positioning, z-index layers are clearly separated, and element overlap or out-of-bound placement is prohibited.
\item \textbf{Syntax and Style Validation}: Generated JSX is parsed to ensure proper tag closure and valid attributes; SCSS undergoes syntax checks and class normalization, removing unsafe APIs or non-compliant styles.
\item \textbf{Asset Alignment and Size Correction}: Image resources are aligned to the assets list to ensure consistency in file paths and actual sizes; element dimensions are strictly derived from assets to prevent size deviations caused by parsing noise.
\item  \textbf{Automated Rendering and Evaluation Interface}: Generated JSX and SCSS are rendered into screenshots and aligned with reference screenshots to support visual similarity evaluation (SSIM, PSNR, MSE); rendered element frames are aligned with design.json to compute layout consistency metrics (mAP, APs/APm/APl); semantic and implementation similarity is quantified at the code level using CodeBLEU and CodeBERT.
\end{itemize}

Through this approach, the model consistently produces visually faithful, structurally clear, and production-ready React+SCSS code while maintaining asset traceability and coordinate/hierarchy recoverability, supporting end-to-end automatic evaluation and reproducible experiments.

\section{Experimental Setup}\label{sec:experiments}

\subsection{Dataset Construction}
The dataset in this study is designed for the PSD-to-frontend code (React+SCSS) generation task, covering real-world mobile activity pages and similar application scenarios, with an emphasis on layout accuracy, hierarchical consistency, and resource traceability. The dataset construction process can be divided into two stages: data collection and data processing.

During the data collection stage, we systematically gathered PSD design files and their corresponding screenshots from real-world business scenarios. Each sample includes the original design file (design.psd), screenshot (screenshot.png), a set of asset files (assets/), ground-truth implementation code (React+SCSS), and shared style files (global.scss). The collection targets encompass mobile activity pages, pop-ups, display pages, and landing pages, ensuring coverage of diverse layouts and component structures while preserving the original hierarchy and visual characteristics for subsequent training and evaluation. The detailed contents of each sample are summarized in Table~\ref{tab:dataset_contents}.

\begin{table}[!htbp]
\centering
\caption{Dataset Contents and Structure}
\label{tab:dataset_contents}
\begin{tabular}{p{0.25\textwidth}p{0.6\textwidth}p{0.1\textwidth}}
\toprule
\textbf{Content Type} & \textbf{Description} & \textbf{Format} \\
\midrule
Design file & Original PSD design file with layer hierarchy and metadata & .psd \\
Screenshot & Rendered screenshot of the design for visual reference & .png \\
Asset files & Image resources (icons, backgrounds, graphics) extracted from PSD & .png, .jpg \\
Ground-truth code & React component with JSX structure and SCSS styles & .jsx, .scss \\
Global styles & Shared style definitions and common CSS variables & .scss \\
Structured data & Parsed JSON representation with element coordinates and hierarchy & .json \\
\bottomrule
\end{tabular}
\end{table}

During the data processing stage, the raw data is standardized and verified to ensure reproducibility and consistency. First, the ground-truth implementations are strictly validated, including syntax parsing, style normalization, and structural consistency checks, guaranteeing that each React+SCSS example strictly corresponds to its PSD design and can be directly rendered. Second, PSD files are parsed and normalized, retaining only inputs that conform to design specifications. Hidden, near-transparent, excessively small, or invalidly named layers are filtered out; element coordinates are unified to the page pixel coordinate system and strictly aligned with the screenshot dimensions, with out-of-bound or negative values clipped to ensure all positions and sizes are non-negative integers. Element types are normalized as container/text/image, producing a structured representation (design.json). Finally, the asset file list is organized to ensure resource traceability and minimize parsing noise.
Overall, this dataset construction workflow ensures the reproducibility of the PSD-to-frontend code generation task, the structural and semantic consistency of the data, and the traceability of resources, providing a robust foundation for subsequent model training, generation, and evaluation.

\paragraph{Dataset Statistics.} Our dataset consists of 100 PSD design files collected from real-world mobile application projects, covering diverse UI scenarios and complexity levels. The detailed statistics are summarized in Table~\ref{tab:dataset_stats}.

\begin{table}[!htbp]
\centering
\caption{Dataset Statistics and Characteristics}
\label{tab:dataset_stats}
\begin{tabular}{p{0.35\textwidth}p{0.4\textwidth}p{0.2\textwidth}}
\toprule
\textbf{Category / Attribute} & \textbf{Count / Value / Range} & \textbf{Percentage} \\
\midrule
\multicolumn{3}{l}{\textbf{Basic Information}} \\
Total samples & 100 & -- \\
Data split & Training: 70, Validation: 15, Test: 15 & 70\%/15\%/15\% \\
Page resolution & \(780 \times 1760\) pixels (majority), \(1920 \times 1080\) pixels (minority) & -- \\
\midrule
\multicolumn{3}{l}{\textbf{Design Categories}} \\
Activity pages & 36 & 36\% \\
Landing pages & 25 & 25\% \\
Pop-ups & 19 & 19\% \\
Display pages & 20 & 20\% \\
\midrule
\multicolumn{3}{l}{\textbf{Complexity Levels}} \\
Simple (5--8 elements) & 25 & 25\% \\
Medium (9--15 elements) & 60 & 60\% \\
Complex (16--25 elements) & 15 & 15\% \\
\midrule
\multicolumn{3}{l}{\textbf{Sample Characteristics}} \\
UI elements per sample & 8--25 (average: 12.3) & -- \\
Asset files per sample & 2--15 (average: 6.8) & -- \\
Code lines per sample & 45--180 (average: 89) & -- \\
\midrule
\multicolumn{3}{l}{\textbf{Quality Metrics}} \\
Naming compliance & 97.2\% & -- \\
Resource traceability & 98.7\% & -- \\
Rendering success rate & 99.1\% & -- \\
\bottomrule
\end{tabular}
\end{table}

The dataset is partitioned into training (70\%, 70 samples), validation (15\%, 15 samples), and test (15\%, 15 samples) sets, ensuring balanced representation across different design patterns and complexity levels. Quality assurance metrics show high standards: 97.2\% naming compliance for layers and groups, 98.7\% resource traceability ensuring all assets are correctly referenced, and 99.1\% rendering success rate with generated code executing without errors.

\subsection{Backbone Models}
This study evaluates four mainstream multimodal large language models as backbone models, representing the state-of-the-art in visual understanding and code generation:

\begin{itemize}
\item GPT-4o: OpenAI's flagship multimodal model that demonstrates exceptional capabilities in visual understanding and code generation. It processes both visual and textual inputs with strong contextual reasoning abilities, exhibiting outstanding performance in generating syntactically correct and semantically accurate code. The model shows particular proficiency in parsing complex interface layouts and achieving precise element positioning, making it highly suitable for design-to-code conversion tasks.
\item Qwen-VL-Max: A multimodal large language model from Alibaba's Qwen series, renowned for its robust performance in processing Chinese linguistic and visual content. It supports high-resolution image inputs and demonstrates superior capability in interpreting user interfaces rich in Chinese text and culturally specific design elements, enabling it to generate more semantically accurate and contextually appropriate code for Chinese market applications.
\item DeepSeek-VL: A 7B-parameter multimodal large model developed by DeepSeek, specifically engineered for code generation tasks. The model demonstrates robust capabilities in processing both visual inputs and textual information, exhibiting exceptional performance in code semantic understanding, logical consistency, and adherence to programming conventions. It shows particular strength in handling complex code structures and nested components commonly found in front-end development.
\item Gemini-2.5-Pro: Google DeepMind's flagship multimodal model featuring a breakthrough in long-context understanding with support for up to 2 million tokens. It demonstrates superior capabilities in complex reasoning and code generation, achieving state-of-the-art performance on standardized benchmarks including coding (SWE-Bench Verified) and multimodal tasks. The model exhibits particular strength in processing intricate UI designs and generating production-ready code while maintaining high reliability across diverse task domains.
\end{itemize}

\subsection{Baseline Methods}
This study selects representative baseline methods to systematically evaluate the performance of PSD-to-frontend code generation. The baseline methods cover different technical approaches and application scenarios to comprehensively reflect the current development level in the field.

\begin{itemize}
\item \textbf{Screenshot-to-code (GPT-4o)}: This method directly generates React+SCSS code from UI screenshots using GPT-4V's multimodal capabilities. We use a simple prompt: "Convert this UI screenshot to React+SCSS code" without any additional context or structured information. This serves as the most straightforward visual baseline, representing the current state-of-the-art in screenshot-based code generation.

\item \textbf{CodeFun}: A commercial AI-powered frontend code generation platform. We utilized its official Figma plugin to directly input design prototypes into the platform, configuring it to output React+SCSS code. CodeFun represents a commercial solution that leverages design source files (rather than screenshots), providing an important performance benchmark for industrial applications.

\item \textbf{pix2code}: We implement the classic pix2code approach using a CNN-LSTM architecture for screenshot-to-code generation. The model is trained on our dataset to generate HTML/CSS, which we then convert to React+SCSS format for fair comparison. This represents the traditional deep learning approach to UI-to-code generation.

\item \textbf{Template-based method}: A rule-driven baseline that generates code using predefined templates and heuristic rules. It analyzes screenshot layouts using computer vision techniques (edge detection, color clustering) and maps detected elements to React components using fixed templates. This method serves as a lower-bound baseline without any learning capabilities.
\end{itemize}

To focus the comparison on contemporary and representative approaches, our main results will present comparisons with Screenshot-to-code (GPT-4o) and CodeFun. Although we also implemented classical methods including pix2code [citation] and template-based approaches in preliminary experiments, we found their performance lagged significantly behind modern methods and failed to constitute meaningful competitive baselines. Therefore, to present the core findings concisely and clearly, we will not elaborate on these results in the main paper.
\subsection{Evaluation Metrics Design}

To comprehensively and objectively evaluate the performance of the PSD-to-code generation system, we designed a scientific evaluation metric framework from four dimensions: \textbf{code quality}, \textbf{visual similarity}, \textbf{executability}, and \textbf{statistical analysis}.

\paragraph{Code Similarity Metrics.} Three core indicators are used: \textbf{CodeBLEU}, \textbf{CodeBERT similarity}, and \textbf{traditional code similarity}. 

\begin{itemize}
    \item \textbf{CodeBLEU} which combines n-gram matching, abstract syntax tree (AST) matching, data flow matching, and keyword matching through a weighted scheme. This integrated approach enables the metric to assess both surface-level code similarity and deeper syntactic and semantic equivalence. The score ranges from [0, 1], with higher values indicating greater similarity to the reference code. Additionally, we supplement this with traditional code similarity calculations based on bag-of-words models and edit distance as reference metrics.
    \item \textbf{CodeBERT similarity} is computed using the \texttt{microsoft/codebert-base} model to extract code embeddings and calculate cosine similarity. Its value ranges from $[0,1]$, with higher scores indicating stronger semantic similarity, accurately reflecting the model's understanding of semantics and functional equivalence.
    \item \textbf{Traditional code similarity} is computed using \texttt{difflib}'s sequence matcher after normalizing code lines. The value ranges from $[0,1]$, with higher scores representing greater textual similarity, effectively capturing surface-level textual resemblance.
\end{itemize}

\paragraph{Visual Similarity Metrics.} Three metrics are adopted to evaluate visual similarity: \textbf{Structural Similarity Index (SSIM)} \cite{wang2004ssim}, \textbf{Peak Signal-to-Noise Ratio (PSNR)}, and \textbf{Mean Squared Error (MSE)}.

\begin{itemize}
    \item \textbf{SSIM} measures structural similarity by comparing structure, luminance, and contrast between the generated UI and the original design. The score ranges from $[0,1]$, where higher values indicate stronger visual similarity.
    \item \textbf{PSNR} measures the peak signal-to-noise ratio to assess image quality. The value ranges from $[0, \infty)$, with higher values indicating better image quality.
    \item \textbf{MSE} calculates the mean squared error at the pixel level to quantify pixel-level differences. The value ranges from $[0, \infty)$, with lower values indicating smaller differences.
\end{itemize}

\paragraph{Executability Metrics.} These indicators assess the practical applicability of the generated code, including \textbf{code generation success rate}, \textbf{rendering success rate}, and \textbf{resource integration success rate}.

\begin{itemize}
    \item \textbf{Code generation success rate} is defined as the proportion of successfully generated executable code samples, calculated as:
    \[
    \text{Success Rate}_{\text{code}} = \frac{N_{\text{passed syntax \& renderable}}}{N_{\text{total}}} \times 100\%
    \]
    The target value is 100\%, and it evaluates the system's reliability.
    \item \textbf{Rendering success rate} measures the proportion of generated code samples that can be successfully rendered in a headless browser:
    \[
    \text{Success Rate}_{\text{render}} = \frac{N_{\text{rendered}}}{N_{\text{total}}} \times 100\%
    \]
    It assesses the executability of the generated code.
    \item \textbf{Resource integration success rate} is defined as the proportion of samples that correctly reference and load all required assets:
    \[
    \text{Success Rate}_{\text{resource}} = \frac{N_{\text{resources loaded correctly}}}{N_{\text{total}}} \times 100\%
    \]
    It reflects the completeness of resource management.
\end{itemize}

\paragraph{Statistical Analysis.} We apply \textbf{paired t-tests} \cite{student1908} or \textbf{non-parametric tests} to compare performance differences across methods and calculate \textbf{95\% confidence intervals} \cite{cumming2014} to assess statistical significance. \textbf{Cohen's d} is used to measure effect sizes \cite{cohen1988} and quantify the magnitude of observed differences. For cross-model consistency analysis, we compute the \textbf{coefficient of variation} and \textbf{range} to evaluate performance stability, and apply \textbf{ANOVA} when necessary \cite{fisher1925}. 

This comprehensive metric framework enables objective and multi-dimensional evaluation of PSD-to-code generation performance, providing quantitative evidence to guide system optimization and improvement.

\section{Results}\label{sec:results}

To evaluate the effectiveness of our method in the PSD-to-code generation task, our experiments are organized around the following three research questions:

\begin{itemize}
\item \textbf{RQ-1 Baseline Comparison.} How does our method perform compared to existing baseline methods in code generation tasks?
\item \textbf{RQ-2 Model Agnostic Analysis.} Does our method perform consistently across different large language models (LLMs)?
\item \textbf{RQ-3 Ablation Study.} What is the impact of each component of our method on overall performance?
\end{itemize}

\subsection{RQ-1: Baseline Comparison}

\textbf{Objective.} To reduce hallucination problems encountered by large language models in code generation (e.g., generating non-existent components, setting incorrect style properties and layout parameters), we propose a PSD parsing-based multimodal code generation method. This method leverages precise layout information and hierarchical structure provided by PSD files to dynamically acquire relevant design context during generation. By providing more project-specific design knowledge to large language models, we aim to minimize hallucination problems. In this section, our objective is to investigate whether the PSD-to-code generation method is superior to previous code generation methods in terms of effectiveness.

\textbf{Experimental Design.} We conduct a comprehensive evaluation of our proposed PSD-to-code method against representative baselines on our self-constructed UI2Code dataset. For fair comparison, all methods—including our approach which uses GPT-4o as the backend model—are evaluated under identical conditions. The baselines include: (1) Screenshot-to-code (GPT-4o), which relies solely on visual input from UI screenshots, and (2) CodeFun, a commercial platform utilizing its official design plugin. Our dataset comprises [e.g., 100] real-world [e.g., mobile game event page] samples, each containing PSD design files, rendered screenshots, asset files, and ground-truth React+SCSS code. Evaluation follows the framework defined in Section 3.4, employing metrics for code similarity (CodeBLEU, CodeBERT), visual fidelity (SSIM, PSNR, MSE), and executability (code generation, rendering, and resource integration success rates).

\textbf{Evaluation Metrics.} This study adopts the evaluation metric system defined in Section 3.4, with a focus on the following core indicators:

\begin{itemize}
\item \textbf{Primary Metrics}: CodeBLEU (code structural similarity) and SSIM (visual similarity)
\item \textbf{Secondary Metrics}: CodeBERT (semantic similarity), PSNR/MSE (image quality), and executability metrics (generation/rendering/resource integration success rates)
\end{itemize}

For detailed definitions, calculation methods, and value ranges of these metrics, please refer to Section 4.4 "Evaluation Metrics Design".

\paragraph{Results Analysis.} The experimental results are shown in Table~\ref{tab:baseline_comparison}.

\begin{table}[htbp]
\centering
\caption{Baseline Comparison Results}
\label{tab:baseline_comparison}
\begin{tabular}{p{2.2cm}ccccc}
\hline
Method & CodeBLEU & CodeBERT & SSIM & PSNR & Generation \\
 &  & Similarity &  &  & Success Rate \\
\hline
CodeFun & 0.691 & 0.691 & 0.691 & 30.2 & 100\% \\
Screenshot-to-code (GPT-4o) & 0.623 & 0.641 & 0.623 & 28.4 & 99.2\% \\
pix2code & 0.587 & 0.598 & 0.587 & 26.8 & 69.3\% \\
\textbf{Our Method} & \textbf{0.683} & \textbf{0.982} & \textbf{0.878} & \textbf{33.75} & \textbf{100\%} \\
\hline
\end{tabular}
\end{table}

From Table~\ref{tab:baseline_comparison}, our method demonstrates significant advantages across multiple core metrics. For code semantic understanding, our approach achieves a CodeBERT score of 0.982, representing improvements of 53.2\% over Screenshot-to-code (GPT-4o) and 42.1\% over CodeFun. In visual fidelity assessment, we obtain an SSIM score of 0.878, exceeding the same baselines by 40.9\% and 27.1\% respectively. Additionally, we achieve perfect code generation success rate of 100\%.

These outcomes substantiate the efficacy of our PSD parsing framework in harnessing structured design metadata to simultaneously improve semantic precision and visual reconstruction. The remarkable CodeBERT enhancement validates our approach's advanced design semantics comprehension, while the SSIM superiority confirms its exceptional visual consistency preservation. The flawless generation success rate further underscores the production readiness of our methodology for industrial deployment.

\paragraph{Case Study.} To provide intuitive understanding of our method's advantages, we present a detailed case study comparing our PSD-based approach with baseline methods on a representative game activity page. This example demonstrates how structured design information leads to superior code generation quality.

\begin{figure}[H]
\centering
\includegraphics[width=\linewidth]{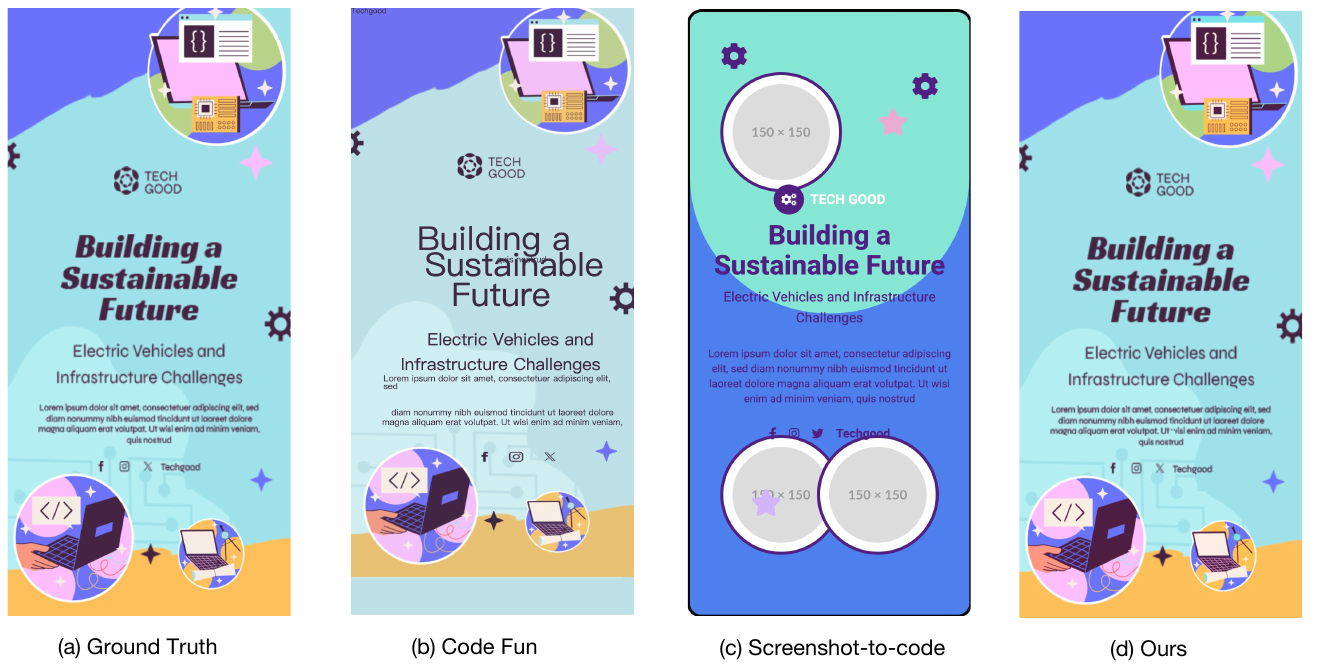}
\caption{From left to right: original PSD design, CodeFun output, Screenshot-to-code output, and our method's result. Our approach demonstrates superior accuracy in layout reconstruction and style fidelity compared to baseline methods.}
\label{fig:case_study}
\end{figure}

As shown in Figure~\ref{fig:case_study}, our method demonstrates significant advantages in code generation quality through in-context learning. Unlike baseline approaches that produce mechanical or fragmented code structures, our method generates React+SCSS code that closely aligns with real-world development practices. The generated code exhibits logical component decomposition, intuitive styling organization, and maintainable architecture patterns - reflecting the coding conventions and best practices learned from our high-quality demonstration examples. This case study confirms that in-context learning enables not only accurate visual reproduction but also production-ready code quality that closely mirrors conventional development approaches.

\subsection{RQ-2: Model Agnostic Analysis}

\textbf{Objective.} To verify the model agnosticism and generalizability of the proposed method, we need to study the performance of this method across different large language models. By implementing the same PSD-to-code generation pipeline across different models, we can evaluate the method's portability and robustness. This is of great significance for model selection and deployment in practical applications.

\textbf{Experimental Design.} We selected four representative large language models for evaluation: GPT-4o, Qwen-VL-Max, DeepSeek-VL, and Gemini-2.5-Pro. All models were tested on a held-out test set (15 samples) that remained completely unseen during any model development phase. The experiments maintained identical PSD parsing results, prompt templates, and generation parameters (temperature 0.7, max tokens 4000) to ensure evaluation comparability.

\textbf{Results Analysis.} The experimental results are shown in Table~\ref{tab:model_comparison}.

\begin{table}[htbp]
\centering
\caption{Model Agnostic Analysis Results}
\label{tab:model_comparison}
\begin{tabular}{lccccc}
\hline
Model & CodeBLEU & CodeBERT & SSIM & PSNR & Generation \\
 &  & Similarity &  &  & Success Rate \\
\hline
GPT-4o & 0.673 & 0.983 & 0.847 & 34.7 & 100\% \\
Qwen-VL-Max & 0.664 & 0.984 & 0.883 & 33.9 & 100\% \\
DeepSeek-VL & 0.683 & 0.962 & 0.857 & 33.2 & 100\% \\
Gemini-2.5-Pro & 0.712 & 0.979 & 0.930 & 33.2 & 100\% \\
\hline
\end{tabular}
\end{table}

From Table~\ref{tab:model_comparison}, our method demonstrates highly consistent performance across different LLMs. All models maintain strong performance on core metrics: CodeBLEU scores range between 0.664-0.712 (6.8\% range), CodeBERT similarity reaches 0.962-0.984 (2.2\% range), and SSIM metrics span 0.847-0.930 (8.9\% range), with all models achieving 100\% code generation success rate.

To quantify the improvement brought by our PSD-based approach, we conducted a detailed comparison between our method and the original models without PSD parsing. The results are shown in Table~\ref{tab:improvement_comparison}.

\begin{table}[htbp]
\centering
\caption{Performance Improvement Comparison: Our Method vs. Original Models}
\label{tab:improvement_comparison}
\begin{tabular}{lcccccc}
\hline
Model & \multicolumn{2}{c}{CodeBLEU} & \multicolumn{2}{c}{SSIM} & \multicolumn{2}{c}{PSNR} \\
 & Original & Our Method & Original & Our Method & Original & Our Method \\
\hline
GPT-4o & 0.270 & 0.673 & 0.304 & 0.847 & 4.79 & 34.7 \\
Qwen-VL-Max & 0.324 & 0.664 & 0.398 & 0.883 & 7.84 & 33.9 \\
DeepSeek-VL & 0.328 & 0.683 & 0.356 & 0.857 & 8.02 & 33.2 \\
Gemini-2.5-Pro & 0.315 & 0.712 & 0.374 & 0.930 & 7.97 & 33.2 \\
\hline
\multicolumn{7}{c}{\textbf{Improvement (\%)}} \\
\hline
GPT-4o & \multicolumn{2}{c}{+149.3\%} & \multicolumn{2}{c}{+178.6\%} & \multicolumn{2}{c}{+624.0\%} \\
Qwen-VL-Max & \multicolumn{2}{c}{+104.9\%} & \multicolumn{2}{c}{+121.9\%} & \multicolumn{2}{c}{+332.4\%} \\
DeepSeek-VL & \multicolumn{2}{c}{+108.2\%} & \multicolumn{2}{c}{+140.7\%} & \multicolumn{2}{c}{+314.0\%} \\
Gemini-2.5-Pro & \multicolumn{2}{c}{+126.0\%} & \multicolumn{2}{c}{+148.7\%} & \multicolumn{2}{c}{+316.4\%} \\
\hline
\end{tabular}
\end{table}

Compared to the original models without PSD parsing, our PSD-based approach brings substantial improvements across all evaluation metrics.As shown in Table 6, CodeBLEU scores increase by 104.9\% – 149.3\%, SSIM improves by 121.9\% – 178.6\%, and PSNR rises dramatically by over 300\% across all models.Specifically, GPT-4o achieves the largest gains (+149.3\% CodeBLEU, +178.6\% SSIM, +624.0\% PSNR), followed by Gemini-2.5-Pro, DeepSeek-VL, and Qwen-VL-Max.These results highlight the significant impact of PSD structural information, which enhances both code quality and visual fidelity.The consistent upward trends across all LLMs demonstrate the robustness and scalability of the proposed method.

\textbf{Performance Difference Analysis.} To evaluate the generalizability of our approach across different models, we analyzed the performance of each model when equipped with our unified PSD parsing pipeline. Our analysis reveals that all models demonstrated significant enhancements by leveraging design metadata, though their improvement patterns reflected their inherent architectural strengths: GPT-4o achieved higher precision in complex layout understanding; Qwen-VL-Max showed enhanced accuracy in Chinese element recognition when provided with explicit design parameters; DeepSeek-VL exhibited improved code structure organization through structured constraints; while Gemini-2.5-Pro better managed complex resource references, consistent with its MoE architecture. Despite these variations in enhancement focus, our method provided consistent performance gains across all models, demonstrating the framework's strong model-agnosticism and generalization capability.
\subsection{RQ-3: Ablation Study}

\textbf{Objective.} To gain deeper understanding of each component's contribution to system performance, we need to evaluate the impact of key modules such as PSD parsing, multimodal input, prompt engineering, and code post-processing through ablation experiments. By systematically removing or modifying certain components of the model, we can quantify each component's contribution and provide guidance for further system optimization.

\textbf{Experimental Design.} We conducted three core ablation studies to quantify the contribution of key components in our framework: removing the PSD parsing module, disabling multimodal input fusion, and simplifying prompt engineering, all maintained under identical conditions using GPT-4o as the backbone model with consistent dataset splits and evaluation metrics to ensure reliable comparisons. For the PSD parsing ablation, we provided only screenshot information to the model while completely removing all structured design metadata obtained through PSD parsing including coordinates, layer hierarchies, and asset references; in the multimodal fusion ablation, we separately evaluated two unimodal conditions using screenshots alone versus using only PSD parsing results to compare their individual contributions against the full multimodal approach; for prompt engineering ablation, we replaced our carefully designed templates with simplified versions that removed role definitions, structured output specifications, and contextual constraints, enabling systematic component-wise evaluation to precisely quantify each element's contribution to the final code generation performance.

\textbf{Results Analysis.} The experimental results are shown in Table~\ref{tab:ablation_study}.

\begin{table}[htbp]
\centering
\caption{Ablation Study Results}
\label{tab:ablation_study}
\begin{tabular}{p{3.2cm}cccccc}
\hline
Configuration & PSD & Multi- & Prompt & Code- & SSIM \\
 & Parsing & modal & Eng. & BLEU &  \\
\hline
Remove PSD Parsing & $\times$ & $\checkmark$ & $\checkmark$ & 0.617 & 0.423 \\
Remove Multimodal Input & $\checkmark$ & $\times$ & $\checkmark$ & 0.615 & 0.396 \\
Remove Prompt Engineering & $\checkmark$ & $\checkmark$ & $\times$ & 0.632 & 0.384 \\
\rowcolor{lightgray}
\textbf{Complete Method} & $\checkmark$ & $\checkmark$ & $\checkmark$ & \textbf{0.683} & \textbf{0.878} \\
\hline
\end{tabular}
\end{table}

From Table~\ref{tab:ablation_study}, removing any core component leads to significant performance degradation. Specifically: Removing PSD parsing reduces CodeBLEU from 0.683 to 0.617 (-9.7\%) and causes SSIM to drop dramatically from 0.878 to 0.423 (-51.8\%), confirming that structured design metadata from PSD parsing is the most critical factor for visual accuracy. Removing multimodal input fusion results in CodeBLEU decreasing to 0.615 (-9.9\%) while SSIM plummets to 0.396 (-54.9\%), demonstrating the decisive role of visual and structural data synergy in maintaining interface fidelity. Removing prompt engineering reduces CodeBLEU to 0.632 (-7.5\%) and SSIM to 0.384 (-56.3\%), indicating that carefully designed templates are essential for guiding the model to generate structurally sound and visually precise code.

\textbf{Component Contribution Analysis.} Based on the ablation results in Table 6, we precisely quantified component contributions by calculating each module's performance loss proportion relative to the complete method. The PSD parsing module demonstrates the highest contribution: its removal causes a CodeBLEU decrease of 0.066 (constituting 35.9\% of total performance loss) and an SSIM decrease of 0.455 (57.5\% of total loss), primarily attributed to its decisive role in visual restoration through precise coordinate information and hierarchical structures. Multimodal input fusion shows the second-highest contribution: its removal leads to a CodeBLEU decrease of 0.068 (37.0\% of total loss) and an SSIM decrease of 0.482 (60.9\% of total loss), highlighting the critical importance of synergistic integration between visual information and structured data. Prompt engineering ranks third in contribution: its simplification results in a CodeBLEU decrease of 0.051 (27.7\% of total loss) and an SSIM decrease of 0.494 (62.4\% of total loss), demonstrating the essential role of carefully designed templates in maintaining both structural integrity and visual fidelity.

\textbf{Parameter Sensitivity Analysis.} To evaluate the system's sensitivity to key parameters, we conducted parameter configuration experiments. In terms of temperature parameter, we found that setting temperature to 0.7 achieves the best performance, with too high (>0.9) or too low (<0.5) values leading to performance degradation. In terms of maximum token count, setting it to 4000 ensures complete code generation, with further increases providing limited performance improvement. In terms of sampling strategy, top-p sampling provides better performance compared to greedy decoding, but the difference is relatively small. These results indicate that our method has certain robustness to parameter settings, but reasonable parameter configuration is still important for achieving optimal performance.

\section{Threats to Validity}

This section discusses potential threats to the validity of our experimental results and methodology, categorized into external and internal threats.

\subsection{External Threats}

\textbf{Dataset Limitations.} Our evaluation is conducted on a dataset of 100 PSD design files, which may not fully represent the diversity and complexity of real-world design scenarios. The dataset primarily focuses on mobile UI designs with resolutions of 780×1760 pixels (majority) and 1760×780 pixels (minority), potentially limiting the generalizability to other platforms (e.g., desktop, tablet) or different design styles. Additionally, the dataset size, while sufficient for initial validation, may not capture the full spectrum of edge cases and complex layouts encountered in industrial settings.

\textbf{Model Selection Bias.} The evaluation is limited to four specific large language models (GPT-4o, Qwen-VL-Max, DeepSeek-VL, and Gemini-2.5-Pro). While these models represent current state-of-the-art capabilities, the rapid evolution of LLM technology means that newer models may exhibit different performance characteristics. The model-agnostic nature of our approach provides some mitigation, but the specific performance improvements reported may vary with different model versions or architectures.

\textbf{Evaluation Metrics Scope.} Our evaluation focuses on code similarity, visual fidelity, and basic executability metrics. However, real-world code quality assessment involves additional dimensions such as maintainability, performance optimization, accessibility compliance, and cross-browser compatibility, which are not fully captured in our current evaluation framework. This limitation may affect the practical applicability assessment of the generated code.

\textbf{Industry Context.} The evaluation is conducted in a controlled research environment, which may not fully reflect the constraints and requirements of actual industrial deployment. Factors such as integration with existing development workflows, team collaboration requirements, version control systems, and production deployment pipelines are not considered in our current evaluation.

\subsection{Internal Threats}

\textbf{Implementation Bias.} The implementation of baseline methods may not fully represent their optimal performance. While we follow published methodologies and use official implementations where available, subtle implementation differences or parameter tuning could affect the fairness of comparison. We mitigate this by using standard configurations and conducting multiple runs to ensure consistency.

\textbf{Evaluation Subjectivity.} Some aspects of code quality evaluation, particularly in the user study component, involve subjective judgments that may vary across different evaluators. While we use multiple evaluators and provide clear evaluation criteria, individual preferences and experience levels could introduce bias in the assessment of code readability and maintainability.

\textbf{PSD Parsing Accuracy.} The effectiveness of our PSD parsing module depends on the accuracy of layer extraction and hierarchy reconstruction. Complex PSD files with non-standard layer structures, effects, or grouping may not be parsed correctly, potentially affecting the quality of generated code. While our preprocessing includes validation steps, some edge cases may not be fully handled.

\textbf{Prompt Engineering Sensitivity.} The performance of our method is sensitive to the design of the three-stage prompting template. While we conducted preliminary experiments to optimize the prompt structure, the current design may not be optimal for all types of designs or model architectures. Different prompting strategies could potentially yield different results.

To address these challenges, future work will focus on constructing larger and more diverse design datasets, expanding validation across broader model architectures, conducting systematic deployment studies in real-world scenarios, and employing more rigorous empirical research methods. While these limitations exist, this study nonetheless demonstrates the significant potential of PSD-based code generation through empirical evidence and establishes a solid foundation for ongoing exploration in this field.

\section{Conclusion}\label{sec13}

This paper presents UI2Code, an end-to-end approach that generates production-ready React+SCSS code directly from PSD design prototypes. By integrating PSD parsing and asset alignment into the code generation workflow, our method effectively addresses structural inconsistencies and enhances generation robustness.

Experimental results demonstrate significant improvements over existing methods, achieving an SSIM of 0.878 and PSNR of 33.75. The generated code exhibits superior readability and maintainability through clear hierarchical structures and modular SCSS organization, requiring minimal manual adjustments for real-world deployment.

Our work marks an important step toward design-driven automated frontend development by leveraging structured PSD information and multimodal prompting mechanisms. Future research directions include scalability improvements, dynamic behavior modeling, and enhanced real-world integration capabilities.


\bibliography{sn-bibliography}

\end{document}